\documentclass[12pt]{article}
\usepackage{authblk}

\usepackage{graphics}
\usepackage{graphicx}
\usepackage{amsmath, amsthm, amssymb,url,authblk}
\usepackage{array}
\usepackage{tikz}
\usepackage{natbib}
\usepackage{algorithm} 
\usepackage{algpseudocode} 
\usepackage{lscape}

\usepackage[english]{babel}
\usepackage[margin=1in]{geometry}
\usepackage{setspace}
\usepackage{makecell}

\usepackage{booktabs,dsfont}
\usepackage{fancyhdr}
\usepackage{subcaption}
\usepackage{endnotes}

\usepackage{pdfpages}

\usepackage{xcolor}

\usepackage{array}
\usepackage{caption}
\usepackage{graphicx}
\usepackage{siunitx}
\usepackage[normalem]{ulem}
\usepackage{colortbl}
\usepackage{multirow}
\usepackage{hhline}
\usepackage{calc}
\usepackage{tabularx}
\usepackage{wrapfig}
\usepackage{adjustbox}
\usepackage[hidelinks]{hyperref}

\usepackage{booktabs}
\usepackage{afterpage}
\usepackage{placeins}
\usepackage{tabularx}
\usepackage{dcolumn}
\usepackage{rotating}
\usepackage{adjustbox}
\usepackage{longtable}
\usepackage{breakcites}
\usepackage[para]{threeparttable}
\usepackage{caption}
\usepackage{wrapfig}
\newcommand*{\email}[1]{%
	\normalsize\href{mailto:#1}{}\par
}
\usepackage{subcaption}
  \usepackage{graphicx}

\usepackage{chngcntr} 
\newcounter{suppfigure}

\newcounter{supptable}
\newenvironment{supptable}[1][]{%
  \refstepcounter{supptable}%
  \begin{table}[#1]}{\end{table}}
  
\usepackage{ltablex,booktabs,ragged2e,caption}

\usepackage{sectsty}

\allsectionsfont{\mdseries\centering}

\usepackage{subcaption,graphicx}

\usepackage{fancyhdr}
  \RequirePackage[T1]{fontenc}
\usepackage{lmodern}[lmdh]

\usepackage{wrapfig}

\begin{document}

\title{ChatGPT and the Labor Market: Unraveling the Effect of AI Discussions on Students' Earnings Expectations\\
}

\author{Samir Huseynov\thanks{302 Comer Hall, Auburn, Al, 36830. Correspondence should be directed to szh0158@auburn.edu. I am grateful to Zahra Murad and Wanqi Liang for their valuable feedback. Any remaining errors are solely my responsibility. } \\ Auburn University}

\date{} 

\maketitle

\begin{abstract}
\noindent This paper investigates the causal impact of negatively and positively toned ChatGPT Artificial Intelligence (AI) discussions on US students' anticipated labor market outcomes. Our findings reveal students reduce their confidence regarding their future earnings prospects after exposure to AI debates, and this effect is more pronounced after reading discussion excerpts with a negative tone. Unlike STEM majors, students in Non-STEM fields show asymmetric and pessimistic belief changes, suggesting that they might feel more vulnerable to emerging AI technologies. Pessimistic belief updates regarding future earnings are also prevalent among non-male students, indicating widespread AI concerns among vulnerable student subgroups. Educators, administrators, and policymakers may regularly engage with students to address their concerns and enhance educational curricula to better prepare them for a future that AI will inevitably shape.
        
\end{abstract}



\thispagestyle{empty}
\clearpage
\newpage
\pagenumbering{arabic}
\section{Introduction}
Since its inception on November 30th, 2022, ChatGPT has rapidly attracted over 100 million users, sparking academic and public debates about how Artificial Intelligence (AI) may shape the future of our species \citep{zarifhonarvar2023economics}.  Many tech companies have joined the race to introduce AI technologies (e.g., Microsoft Bing AI, Google Bard, Amazon Codewhisperer, etc.) and built-in products that can profoundly transform the way people live, learn, spend their leisure time, and work.  

Research shows that most firms have adopted at least one AI solution, with those incorporating ChatGPT in daily operations witnessing improved valuations \citep{eisfeldt2023generative}.  These technologies are projected to add approximately 13 trillion USD to the global economy by 2030 \citep{zarifhonarvar2023economics}. Additionally, AI is likely to reshape the labor market by automating cognitively demanding tasks \citep{stephany2021one}. Even in its very preliminary stage of development, Large Language Models (LLM) have already affected 19\% of jobs, with at least 50\% of their tasks being exposed to automation \citep{eloundou2023gpts}. 

Historically, technology-induced unemployment and wage disparities have disrupted low and medium-skilled jobs \citep{acemoglu2020robots, brynjolfsson2018can}. However, technologies like ChatGPT are predicted to impact high-skilled professions, potentially leading to unstable labor market outcomes \citep{eloundou2023gpts, chen2023future, lou2023gpts}. Signs of this shift have emerged as leading tech firms increasingly adopt AI solutions, with reports of hiring freezes and potential layoffs \citep{businessinsider2023ai}.

Natural language AI models, particularly ChatGPT, pose significant challenges for students' career prospects.  ChatGPT has demonstrated superior performance compared to an average student in MBA courses, Bar exams, medical licensing tests, and other cognitively demanding tasks \citep{campello2023new, katz2023gpt,dailymail2023chatgpt}. Today's students face the possibility that AI may partially or entirely overtake their anticipated jobs upon graduation. This challenging scenario could affect projected salaries, pushing students to drop out or switch to ``safer'' majors to secure future earnings \citep{wiswall2015college}. However, it is also possible that AI technologies could potentially enhance future workers' productivity and earning potential \citep{acemoglu2020robots}. They can upskill the workforce, expanding expertise and revitalizing white-collar labor \citep{acemoglu2018low, technologyreview2023}. The discourse on AI, both optimistic and pessimistic, could shape students' educational choices, leading to career-defining decisions. Understanding how public conversations around emerging technologies affect students' outlooks and anticipated earnings is both academically and policy-wise significant.\footnote{The inspiration for this article came from the author's several conversations with his undergraduate students during the Spring 2023 semester. The students' concerns about future job opportunities were recurring in those discussions.} This paper investigates the impact of the ongoing debate about ChatGPT and other natural language AI models on the expected labor market outcomes of university students in the United States. Past research indicates that news media can substantially influence individual expectations, triggering aggregate behavioral and macroeconomic changes \citep{debacker2017expectations,van1989economic}. A detailed exploration of this issue will shed light on how AI discussions in media can shape individual outlooks.

This study used an online experiment with US students to assess how positive and negative public debates about AI affect their anticipated labor market earnings. Using a recent MIT Technology Review article discussing the potential impacts of ChatGPT and related AI tools on the labor force \citep{technologyreview2023}, we developed optimistic (\textit{GoodNews}) and pessimistic (\textit{BadNews}) information-nudge scenarios.

Initially, we elicited students' existing beliefs about their chances of ranking in the top-50\% percentile for post-graduation earnings. Students were then randomly assigned to either the GoodNews or BadNews conditions and presented with corresponding excerpts from the article. Afterward, we collected their revised beliefs, revealing shifts in their views on their chances of ranking in the top-50\% percentile post-graduation. They also expressed their prior and posterior beliefs about a median student with the same major's chances of being in the top-50\% percentile post-graduation. Lastly, students reported their expected annual earnings before and after exposure to the information nudge.

MIT Technology Review holds a 76\% \textit{Factual Grade} according to a fact-checking outlet, reflecting the magazine's accuracy and credibility \citep{technologyreview2023}. Students were provided with this Factual Grade along with the information nudge pieces. This design aspect allowed us to integrate a signal-to-noise ratio into our analyses and formulate theoretical Bayesian posteriors for comparison with students' actual posterior beliefs.

Our findings reveal that exposure to both positive and negative AI ChatGPT discussions leads students to revise down their beliefs about ranking in the top-50\% percentile of post-graduation earnings. The BadNews treatment, however, induces a more significant revision than the GoodNews condition. Interestingly, neither condition influences students' reported expected earnings. This could be due to students' uncertainty about translating current AI developments into potential labor market compensation changes. Furthermore, only the BadNews condition led students to decrease their assessed probability that a median student with the same major will rank in the top-50\% percentile post-graduation; this was not observed in the GoodNews treatment.

Our Bayesian belief updating analysis reveals that students revise their beliefs conservatively after receiving information nudges, deviating from Bayesian benchmarks. Non-STEM majors show asymmetrical and pessimistic belief updating compared to STEM majors, reacting more strongly to negatively framed AI ChatGPT discussions. This suggests that Non-STEM students might feel more threatened by AI advancements. Notably, we observed gender differences in belief updating: non-male students respond more to negative news nudges, aligning with prior research on gender differences in optimism \citep{ayala2014resilience}. Our findings also suggest that students, irrespective of high or low GPA, tend to be more pessimistic about their future earnings potential amid AI-driven labor market shifts.

This research provides initial insights into how recent AI technologies and the ensuing public debate on potential labor market transformations may affect student expectations. Regular engagement with students by educators and policymakers could address their concerns and better assess the potential ramifications of AI technologies on the labor market. Some countries have established regulatory institutions to effectively manage AI-related issues and leverage arising opportunities \citep{halaweh2018artificial}. Society might benefit from extensive research identifying the majors most susceptible to these technological advancements. Higher education institutions could also proactively enrich their core curricula with new skills and knowledge, preparing students for an AI-influenced future. 

\section{Study Procedures and Sample Features}

We conducted our online experiment with students using the Prolific.co crowd-sourcing platform. \citet{dong2023toward} show that field studies using crowd-sourcing platforms are better positioned to reliably represent population-level attitudes towards AI. We restricted our target sample to current students living in the United States and aged 18 or older. We compiled our survey in Qualtrics, and student subjects were compensated with \$2.00 for their participation, corresponding to an hourly rate of \$30.00. The median time for completing the experiment was approximately four minutes.  The study was reviewed and approved by the Auburn University IRB board (IRB: 23-216 EX 2304).\footnote{This study was also registered at a public repository: https://www.socialscienceregistry.org/trials/11364}

Table S1 in Online Appendix presents power calculations and basic sample statistics of our data for 716 subjects, where GoodNews and BadNews experimental conditions have 356 and 360 observations, respectively. A total of 31 observations were removed from the initial sample of 747 participants due to the reporting of zero expected income and/or zero values for both priors and posteriors. Approximately 70\% of our sample is composed of undergraduate students. Nearly half of our subjects have never used ChatGPT or only used it a few times. The average GPA of our sample is close to 3.50, suggesting that our sample primarily consists of students with high GPAs. The distribution between STEM and Non-STEM majors stands at 35\% and 65\%, respectively. The data in Table S1 further confirms that our treatment randomization was successful.


The experiment began with a consent form providing general information about the study. After giving consent, subjects reported their current educational level and major, which later were used to identify STEM or Non-STEM students \citep{ICE2023}. Then participants were shown a brief statement: \textit{``Forbes reports that, based on data from the National Center for Education Statistics, the median starting salary for college graduates is \$59,600 per year.''} Next, we elicited subjects' assessed probability that their starting annual earnings would be above \$59,600 after graduation and the probability that a median student with the same major would have a starting salary above \$59,600 post-graduation. Finally, subjects indicated their anticipated starting salaries. 

During the treatment stage, we provided subjects with a brief overview of how ChatGPT has generated discussions about the potential impact of AI on the labor market, tailoring it to STEM and Non-STEM majors.  This design feature aimed at increasing the ego-relevance of our treatment conditions, which has shown to be important in belief formations and confidence \citep{drobner2022rationalization}. 



We subsequently presented the GoodNews or BadNews information nudge pieces. After reading the treatment information nudge pieces, subjects reported feelings about their earnings prospects and the economy. We also asked participants to report their feelings on the future earnings prospects of students with the same major. These measures serve as manipulation checks identifying the potential channels for the experimental treatments.  In the final stage, we elicited posterior beliefs using the questions we used to quantify the priors. The study concluded with a brief demographic survey and attention check measures.

\section{Estimation Methodology}
Our primary outcome measures of interest include the change in the assessed individual probability of being in the top-50\% percentile of the earnings distribution ($Own$ $Prob_{post}/Prob_{pre}$), the change in expected starting salary levels ($Earning_{post}/Earning_{pre}$), and the change in the perceived probability that others (with the same major) will be in the top-50\% percentile of the earnings distribution ($Others$ $Prob_{post}/Prob_{pre}$). We construct these measures using prior (elicited before providing information nudge pieces) and posterior beliefs (elicited after providing information nudge pieces). We estimate the model specification to examine the impact of our treatment conditions on belief changes as follows:

\begin{equation}
   \Delta_{i} = \alpha_{0}+\alpha_{1}T_{i}+\alpha_{2}*\bold{\Gamma_{i}}+\epsilon_{i}
\end{equation}

where, $\Delta_{i}$ represents the three individual belief change measures constructed for subject $i$; $T_{i}$ is a binary variable that equals one for the GoodNews treatment condition; the vector $\bold{\Gamma_{i}}$ comprises individual demographic measures, including $STEM$ and $Male$, which are binary measures showing if individual $i$ studies in a STEM major and if they identify with being male, respectively; $GPA$ refers to the reported individual Grade Point Averages, while $AdjIncome$ indicates the students' 2022 pre-tax family income, divided by the square root of family size. The term $\epsilon_{i}$ in Equation 1 represents individual idiosyncratic errors. 

\subsection{Bayesian Belief Formation Framework}
We built on the recent advancements in Bayesian belief elicitation literature to assess how our study participants' belief updates after the GoodNews and BadNews experimental treatments align with theoretical posteriors \citep{mobius2022managing, barron2021belief, coutts2019good, drobner2022motivated}. We then identify deviations from the Bayesian benchmark and link them with observable individual characteristics.


We closely follow the framework constructed by \citet{coutts2019good} and \citet{barron2021belief}. Individual $i$ faces two possible future states, $s\in\{G, B\}$. Nature will select one of these states to realize. Individual $i$ forms their prior belief $\pi_{G}$ that $s=G$ in the future.  Consequently, the prior belief of $s=B$ is $\pi_{B}$, with $\pi_{G}+\pi_{B}=1$. Individual $i$ receives a signal, $z\in\{g,b\}$, indicating which state nature will select. This signal is noisy, and $p(g|G)=p(b|B)=q$. In our study, $q$ is 0.76, representing the Factual Grade of the MIT Technology Review. We employ the modeling specification suggested by \citet{mobius2022managing}:

\begin{equation}
  logit(\pi_{G,posterior}) = \delta logit(\pi_{G,prior})+\beta_{G}log(\frac{q}{1-q})*I(z=g)+\beta_{B}log(\frac{1-q}{q})*I(z=b)
\end{equation}

where,$\pi_{G, prior}$ and $\pi_{G, posterior}$ respectively represent the prior and posterior probabilities of being in the top-50\% of the earnings distribution after graduation. Then we use OLS regression and estimate the following specification with individual $\rho_{i}$ errors:

\begin{equation}
  logit(\pi_{G,posterior,i}) = \delta logit(\pi_{G,prior,i})+\beta_{G}log(\frac{q}{1-q})*I(z=g_{i})+\beta_{B}log(\frac{1-q}{q})*I(z=b_{i})+\rho_{i}
\end{equation}

It must be noted that Equation 3 is estimated without the constant, as $I(z=g_{i})+I(z=b_{i})=1$. Moreover, we restrict our sample to \textit{correct} belief updates.\footnote{The belief updating is \textit{correct} if $\pi_{posterior,i}\geq\pi_{prior,i}$ when $z=g_{i}$ or $\pi_{posterior,i}\leq\pi_{prior,i}$ when $z=b_{i}$. We also deal with boundary beliefs $1$ and $0$ by replacing them with $0.99999$ and $0.00001$, respectively.}

\subsection{Interpretation of Bayesian Parameter Values}

Estimating model parameters allows us to identify if participants adhere to the Bayesian model framework in their belief updates \citep{coutts2019good, barron2021belief}. We summarize the possible parameter values and their interpretation as follows:

\begin{minipage}{\linewidth}
\medskip
\medskip
$-$ Bayesian updating if $\delta=1$, $\beta_{G}=1$, and $\beta_{B}=1$.\\
$-$ Conservative updating if $\beta_{G}<1$ or $\beta_{B}<1$.\\
$-$ Overresponding if $\beta_{G}>1$ or $\beta_{B}>1$.\\
$-$ Asymmetric and Optimistic updating if $\beta_{G}>\beta_{B}$.\\
$-$ Asymmetric and Pessimistic updating if $\beta_{G}<\beta_{B}$.
\medskip
\medskip
\end{minipage}

\section{Results}

We begin our discussion of the study's findings by examining responses to our manipulation check questions. Table S2 presents the mean values of reported sentiments in both the GoodNews and BadNews treatments. On average, we find that participants are only optimistic about their own future earnings prospects. The GoodNews treatment notably elevates the average reported sentiment regarding personal earnings prospects by 0.64 $(p-value<0.05)$ compared to the BadNews treatment. However, we observe no statistically significant divergence in reported sentiments about others' earnings prospects and the economy across the study conditions.

These findings suggest that exposing students to optimistically framed AI ChatGPT debates enhances their optimism about future earnings prospects. This insight also hints at the role of \textit{optimism} and \textit{pessimism} as potential causal channels influencing students' educational and subsequent career decisions when facing AI-influenced labor market uncertainties.

 \setcounter{figure}{1}
 \begin{figure}

\centering
\includegraphics[width=1\textwidth]{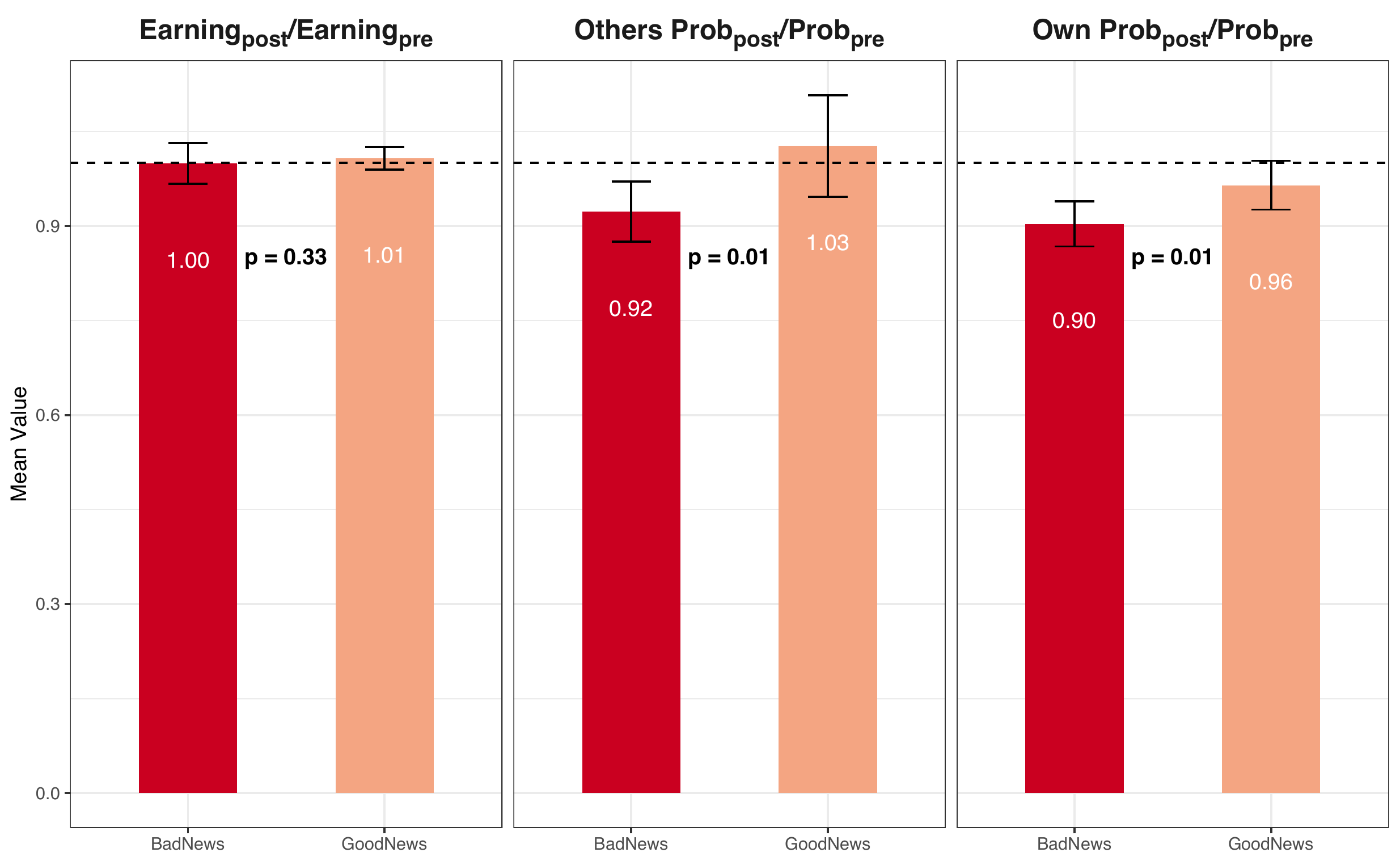}

\medskip 
\begin{minipage}{0.99\textwidth} 
{\footnotesize Note: Elicited individual beliefs on the probability of being in the top-50\% annual earning percentile after graduation, both before and after the treatments: $Own$ $Prob_{pre}$ and $Prob_{post}$. Elicited individual beliefs on the probability that a median student with the same major will be in the top-50\% annual earning percentile after graduation, both before and after the treatments: $Others$ $Prob_{pre}$ and $Prob_{post}$.  Elicited expected annual individual earnings, both before and after the treatments: $Earning_{pre}$ and $Earning_{post}$. The dashed line represents $Mean Value = 1$. T-test p-values are reported. \par}
\end{minipage}

	\caption{\label{fig_1} \textbf{Figure 1}: Treatments and Expected Labor Market Outcomes}
\medskip
\vspace{4pt}
\end{figure}

\medskip
\noindent \textbf{\textit{Result 1: The BadNews information nudge reduces the average value of reported individual probabilities about being in the top-50\% of the earnings distribution after graduation.}}

Figure 1 illustrates the mean value differences in $Own$ and $Others$ $Prob_{post}/Prob_{pre}$, along with $Earning_{post}/Earning_{pre}$ measures, contrasting the GoodNews and BadNews experimental conditions. In both conditions, students reduce their self-assessed probability of being in the top-50\% percentile of post-graduation earnings distributions. Interestingly, the extent of this downward adjustment is significantly greater $(p-value=0.01)$ in the BadNews condition compared to the GoodNews treatment. Our findings suggest that discussions focused on the potential impact of AI technologies on the labor market may erode students' confidence in their future earnings prospects. Furthermore, this effect is more pronounced when AI debates are negatively framed. 

In the GoodNews condition, students' estimated probabilities regarding a median student's (with the same major) earnings prospects remain unaltered. However, in the BadNews treatment, students lower $(p=0.01)$ posterior probabilities. Remarkably, students' estimations of their future earnings levels remain steady, unswayed by the influence of the GoodNews or BadNews information nudges. The result showing that students tend to lower their confidence levels (i.e., assessed probabilities) but not their expected starting salaries could indicate that participants are uncertain about how the changes in the labor market caused by AI will affect their anticipated earnings.

\begin{table}[!htbp] \centering 
\medskip
\medskip
 \caption{\label{table_3}  \textbf{Table 3}: The Impact of AI ChatGPT Discussions on Expected Labor Market Outcomes}
\centering
\scalebox{0.9}{\begin{tabular}{@{\extracolsep{5pt}}lcccccc} 
\\[-1.8ex]\hline 
 & \multicolumn{6}{c}{\textit{Dependent variable:}} \\ 

\\[-1.8ex] & \multicolumn{6}{c}{ } \\ 
 & \multicolumn{2}{c}{\boldmath{$Own$ $Prob_{post}/Prob_{pre}$}} & \multicolumn{2}{c}{\boldmath{$Others$ $Prob_{post}/Prob_{pre}$}} & \multicolumn{2}{c}{\boldmath{$Earning_{post}/Earning_{pre}$}} \\
\cline{2-3} \cline{4-5} \cline{6-7}

\\[-1.8ex] & (1) & (2) & (3) & (4) & (5) & (6) )\\ 

 GoodNews & 0.06$^{**}$ & 0.06$^{**}$ & 0.10$^{**}$ & 0.10$^{**}$ & 0.01 & 0.01 \\ 
  & (0.03) & (0.03) & (0.05) & (0.05) & (0.02) & (0.02) \\ 
 
 STEM &  & $-$0.01 &  & 0.004 &  & $-$0.03$^{*}$ \\ 
  &  & (0.03) &  & (0.03) &  & (0.02) \\ 
 
 Male &  & 0.03 &  & $-$0.02 &  & 0.01 \\ 
  &  & (0.03) &  & (0.04) &  & (0.02) \\ 

 GPA &  & 0.06$^{*}$ &  & 0.12$^{***}$ &  & $-$0.01 \\ 
  &  & (0.03) &  & (0.05) &  & (0.02) \\ 
 
 AdjIncome &  & 0.004 &  & $-$0.001 &  & 0.004 \\ 
  &  & (0.003) &  & (0.01) &  & (0.004) \\ 
 
 Constant & 0.90$^{***}$ & 0.68$^{***}$ & 0.92$^{***}$ & 0.52$^{***}$ & 1.00$^{***}$ & 1.02$^{***}$ \\ 
  & (0.02) & (0.10) & (0.02) & (0.14) & (0.02) & (0.07) \\ 

 \hline

 N & 716 & 716 & 716 & 716 & 716 & 716   \\ 

\hline \\[-1.8ex] 

\end{tabular}  }
\medskip
\begin{minipage}{0.93\textwidth} 

{\footnotesize  Note:   OLS regression results with HC1 robust standard errors are reported. {$^{*}$p$<$0.1; $^{**}$p$<$0.05; $^{***}$p$<$0.01} }
     \end{minipage}
     
\end{table} 

Table 1 expands on the analyses in Figure 1, using regression analyses with controls. We confirm our findings drawn from Figure 1. Based on Table 1 Column 1, the GoodNews experimental condition increases the posterior beliefs about \textit{Own} probability of being in the top-50\% of the earnings distribution by 6 percentage points (p.p.). This result is robust to the inclusion of relevant control variables into the regression specification.  Table 1 Columns 3 and 4 show that the GoodNews treatment increases participants' confidence about \textit{Others'} earning prospects. Interestingly, we also find that students with higher GPAs do not change their confidence in their earnings prospects after reading information nudge excerpts about AI.  

Our analyses also reaffirm the finding from Figure 1 that our treatment conditions do not significantly alter expected earnings levels.  We observe that being in a STEM field is not associated with any expectation changes regarding one's starting salary levels. We do not find any significant effect of gender and family income on our outcome measures.

\begin{table}[!htbp] \centering 
\medskip
\medskip
 \caption{\label{table_4}  \textbf{Table 4}: The Impact of AI ChatGPT Discussions on Belief Updating}
\centering
\scalebox{0.9}{\begin{tabular}{@{\extracolsep{5pt}}lccc} 
\\[-1.8ex]\hline 
 & \multicolumn{3}{c}{\textit{Dependent variable: Logit Belief}} \\ 

\\[-1.8ex] & All & STEM & Non-STEM  \\ 

  \boldmath{$\delta$}  & 0.93$^{***}$ & 0.89$^{***}$ & 0.95$^{***}$ \\ 
  & (0.03) & (0.05) & (0.04) \\ 
  & & & \\ 
 \boldmath{$\beta_{GoodNews}$} & 0.29$^{***}$ & 0.38$^{***}$ & 0.27$^{***}$ \\ 
  & (0.06) & (0.13) & (0.07) \\ 
  & & & \\ 
 \boldmath{$\beta_{BadNews}$}  & 0.74$^{***}$ & 0.49$^{***}$ & 0.86$^{***}$ \\ 
  & (0.10) & (0.09) & (0.14) \\ 
  
   \hline
   
 $ \mathds{P} $ ($\delta=1$) & 0.02 & 0.04 & 0.20 \\ 
 $ \mathds{P} $ ($\beta_{GoodNews}=1$) & 0.00 & 0.00 & 0.00 \\ 
 $ \mathds{P} $ ($\beta_{BadNews}=1$) & 0.01 & 0.00 & 0.31 \\ 
 $ \mathds{P} $ ($\beta_{GoodNews}=\beta_{BadNews}$) & 0.00 & 0.19 & 0.00 \\ 

 N & 476 & 162 & 314   \\ 

\hline \\[-1.8ex] 

\end{tabular}}
\medskip
\begin{minipage}{0.83\textwidth} 

{\footnotesize  Note: Analyses of the impact of logit priors on logit posterior beliefs.  The estimated $\beta_{GoodNews}$ and $\beta_{BadNews}$ show the magnitude of belief updates after treatment conditions.  OLS regression results with HC1 robust standard errors are reported. \\
{$^{*}$p$<$0.1; $^{**}$p$<$0.05; $^{***}$p$<$0.01} }
     \end{minipage}
     
\end{table}

\noindent \textbf{\textit{Result 2:  In the BadNews treatment, students exhibit a disproportionate reaction to negatively toned discussions about AI, showing asymmetric and pessimistic belief updating.}}

Table 2 presents the estimation of Equation 3. In the column labeled ``All,'' the estimated value of $\delta$ is $0.93$, which is statistically different from one $(p-value=0.02)$. Additionally, the condition $\beta_{GoodNews}\neq \beta_{BadNews}$ holds true $(p-value<0.01)$, indicating that study participants do not follow the Bayesian belief updating benchmarks.

The column ``All'' in Table 2 further reveals that the estimated values of $\beta_{GoodNews}$ and $\beta_{BadNews}$ are less than $1$. This implies a conservative belief updating pattern among students. The pessimistic belief updating is confirmed by $\beta_{GoodNews}<\beta_{BadNews}$. Thus, we conclude that the study participants overreacted to the information nudge piece in the BadNews condition, demonstrating asymmetric and pessimistic belief updates.

Table 2 Columns ``STEM'' and ``Non-STEM'' show that students with Non-STEM majors are susceptible to asymmetric and pessimistic belief updating, while STEM majors show the same reaction level to both negatively and positively toned AI discussions. This finding suggests that Non-STEM majors are disproportionally more concerned about the possible adverse effects of ChatGPT and other AI technologies on future labor market earnings prospects.

 \setcounter{figure}{1}
 \begin{figure}

\centering
\includegraphics[width=1\textwidth]{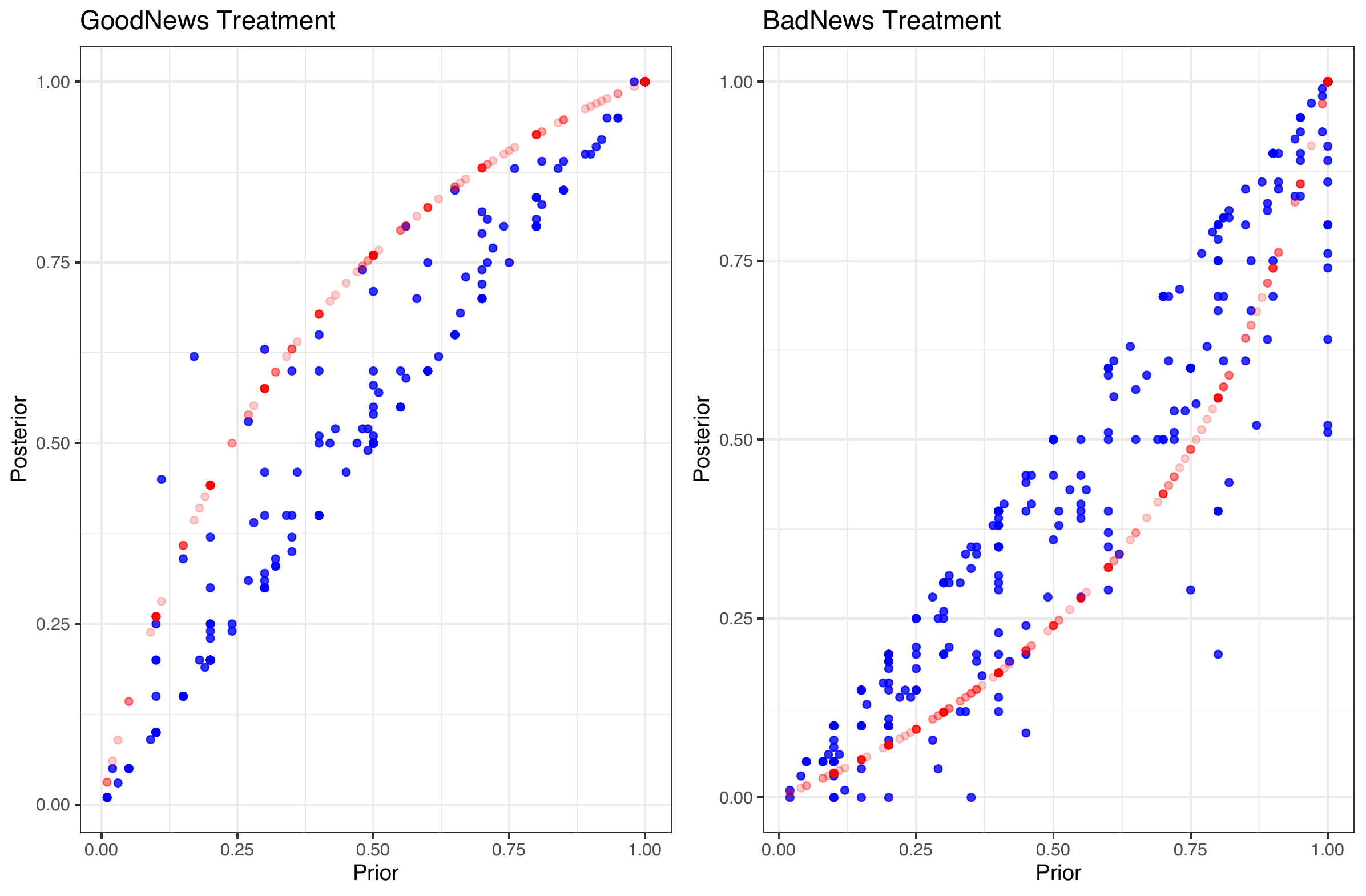}

\medskip 
\begin{minipage}{0.99\textwidth} 
{\footnotesize Note: Blue scatter plot dots represent elicited individual Priors are posteriors about the probability of being in the top-50\% annual earning percentile after graduation. The red scatter plot dots show the theoretical Bayesian posteriors for elicited individual priors.  \par}
\end{minipage}

	\caption{\label{fig_2} \textbf{Figure 2}: Theoretical and Observed Posteriors  }
\medskip
\vspace{4pt}
\end{figure}

In Figure 2, to construct theoretical Bayesian benchmarks (illustrated in red), we utilize elicited priors on the \textit{Own} probability of being in the top 50\% of the earnings distribution upon graduation. These benchmarks are then juxtaposed with observed posteriors (depicted in blue). Our findings from Table 2 are reaffirmed, demonstrating that the actual posteriors tend to be of lower magnitude compared to the theoretical ones. Furthermore, Figure 2 reveals that the BadNews treatment condition yields larger posterior values than the GoodNews condition, further highlighting the presence of asymmetric and pessimistic shifts in belief. Our results confirm the findings of previous studies showing negative news coverage induces widespread reductions in the confidence of economic agents, potentially affecting macroeconomic behavior \citep{hollanders2011influence}. 

\noindent \textbf{\textit{Result 3:  We find gender differences in belief updating. Non-Male students show asymmetric and pessimistic belief changes after the information nudge discussions.}}

\begin{table}[!htbp] \centering 
\medskip
\medskip
 \caption{\label{table_5}  \textbf{Table 5}: The Impact of AI ChatGPT Discussions on Belief Updating \\  across subsamples}
\centering
\scalebox{0.9}{\begin{tabular}{@{\extracolsep{5pt}}lcccc} 
\\[-1.8ex]\hline 
 & \multicolumn{4}{c}{\textit{Dependent variable: Logit Belief}} \\ 

\\[-1.8ex]  & Low-GPA & High-GPA & Male & Non-Male  \\ 

\boldmath{$\delta$} & 0.90$^{***}$ & 0.97$^{***}$ & 0.90$^{***}$ & 1.00$^{***}$ \\ 
  & (0.05) & (0.04) & (0.04) & (0.04) \\ 
  & & & & \\ 
 \boldmath{$\beta_{GoodNews}$} & 0.29$^{***}$ & 0.29$^{***}$ & 0.43$^{***}$ & 0.23$^{***}$ \\ 
  & (0.08) & (0.09) & (0.13) & (0.06) \\ 
  & & & & \\ 
 \boldmath{$\beta_{BadNews}$}  & 0.74$^{***}$ & 0.74$^{***}$ & 0.74$^{***}$ & 0.68$^{***}$ \\ 
  & (0.13) & (0.16) & (0.14) & (0.14) \\ 
  & & & & \\ 

   \hline
   
 $ \mathds{P} $ ($\delta=1$) & 0.03 & 0.33 & 0.02 & 0.99 \\ 
 $ \mathds{P} $ ($\beta_{GoodNews}=1$) & 0.00 & 0.00 & 0.00 & 0.00 \\ 
 $ \mathds{P} $ ($\beta_{BadNews}=1$) & 0.05 & 0.09 & 0.06 & 0.02 \\ 
 $ \mathds{P} $ ($\beta_{GoodNews}=\beta_{BadNews}$) & 0.00 & 0.01 & 0.07 & 0.00 \\ 
 N & 260 & 216 & 236 & 240   \\ 

\hline \\[-1.8ex] 

\end{tabular}  }
\medskip
\begin{minipage}{0.83\textwidth} 

{\footnotesize     OLS regression results with HC1 robust standard errors are reported. \\
{$^{*}$p$<$0.1; $^{**}$p$<$0.05; $^{***}$p$<$0.01} }
     \end{minipage}
     
\end{table}

Table 3 extends our belief-updating analyses using Equation 3, focusing on gender and GPA subgroups. The results reveal that male students exhibit a symmetric yet conservative belief updating when exposed to positive and negative news nudges. However, non-male students appear to react more to negative news.  Our results also indicate both Low- and High-GPA subgroups (when our study sample is divided using a median-point-split approach) show asymmetric belief updating with a higher magnitude of belief changes when faced with negatively-toned news excerpts. These findings lead us to conclude that irrespective of GPA level, students tend to exhibit disproportionately pessimistic reactions to negatively framed ChatGPT discussions. This indicates that AI concerns seem to loom larger across all student subgroups.

\section{Conclusions}

ChatGPT's inception started a new era in human history, signaling that non-human intelligence, through the progression of AI technologies, will shape our future. The economy, and by extension, the labor market, are already exhibiting signs of this inevitable reality.

Modern AI tools uniquely target white-collar professions by automating essential work tasks. This shift is novel, as historically, technological advancements mainly disrupted low- to medium-skilled jobs. Today's students may be more susceptible to these AI-driven transformations, with the expectation that ChatGPT and similar AI tools could replace human labor soon, potentially rendering some academic majors partially or entirely obsolete.

As these substantial changes unfold, public discourse on AI and our future exhibits contrasting views. Leading economists and AI researchers offer both optimistic and pessimistic projections for economic and labor market growth, and wage changes. The tone of these debates may sway current students' educational and career choices. This paper investigates the causal influence of both negatively and positively framed AI discussions on US students' expected labor market outcomes.

Our findings show that students' confidence in their future earnings prospects diminishes upon exposure to AI debates, with a more pronounced effect when encountering negatively framed information. Our Bayesian belief updating analyses also reveal that students exhibit asymmetric and pessimistic belief shifts after exposure to negatively framed ChatGPT discussions. Non-male students demonstrate increased concern following discouraging AI debates, indicating heightened vulnerability to this emerging technology. Pessimistic updates to beliefs about their future earnings prospects are prevalent across both low and high GPA levels, suggesting universal AI concerns among students. We recommend that educators, higher education administrators, and policymakers regularly engage with students to understand their concerns and implement necessary changes to core curricula, better preparing them for a future with Artificial Intelligence.

\FloatBarrier

\newpage

\FloatBarrier

\bibliographystyle{apalike}
\bibliography{ref.bib}

\FloatBarrier
\newpage
\pagestyle{fancy}
\fancyhf{}
\centering\section*{\bfseries{Online Appendix}}
\vspace{1cm}

\section*{ChatGPT and the Labor Market: Unraveling the Effect
of AI Discussions on Students’ Earnings Expectations}
Samir Huseynov

Auburn University

\newpage
\FloatBarrier
 
\begin{supptable}

\caption{\label{table_1}  \textbf{Table 1}: Sample Statistics}
\centering

\setlength{\extrarowheight}{5pt} 
	\scalebox{0.79}{
\begin{tabular}{@{\extracolsep{15pt}}lcc|c} 
\addlinespace

\hline

  &  \textit{BadNews}  &  \textit{GoodNews} & Adj. \textit{P-value}  \\ 
    &    N=360 & N=356 &\\

Family Income (USD) & 46,945 (33,920) & 43,755 (32,270)   & 0.52\\
STEM & 129  (36\%) & 126  (35\%)  & 0.99\\
GPA & 3.48 (0.43) & 3.49 (0.50)  & 0.52\\
Undergraduate & 263  (73\%) & 250  (70\%)  & 0.52\\
Male & 195  (54\%) & 156  (44\%)  & 0.06\\
GPTDaily & 5  (1.4\%) & 13  (3.7\%)  & 0.29\\
GPTRegularly& 33  (9.2\%) & 39  (11\%)  & 0.52\\
GPTFrequently& 64  (18\%) & 51  (14\%)  & 0.52\\
GPTOccasionally & 60  (17\%) & 70  (20\%)  & 0.52\\
GPTRarely & 100  (28\%) & 86  (24\%)  & 0.52\\
GPTNever & 98  (27\%) & 97  (27\%)  & 0.99\\
\hline
\end{tabular}}

\medskip
\begin{minipage}{0.89\textwidth} 

{\small     Note: Mean (Std. Dev) or N (Proportions) are reported. We conducted the Wilcoxon rank sum and Pearson's Chi-squared tests to detect differences between treatment conditions for continuous and categorical measures, respectively. P-values are corrected with the Benjamini \& Hochberg method to account for multiple testing. \\
Rarely $-$ a few times a year; Occasionally $-$ once a month or less, Frequently $-$ a few times a month, Regularly $-$ once a week or more.}
     \end{minipage}

\medskip
\medskip
\end{supptable}
\medskip
\medskip

\begin{table}
\caption{\textbf{Table 2}: Manipulation Checks for Experimental Treatments}
\label{table_2}
\centering
\setlength{\extrarowheight}{5pt} 
\scalebox{0.89}{
\begin{tabular}{lccc}
\hline
 & \textit{Difference} &  \textit{BadNews}  &  \textit{GoodNews}   \\ 
\hline
Feelings about own earnings prospects & 0.64$^{**}$ & 1.65 (0.27) & 2.29 (0.25)  \\
Feelings about others' earnings prospects &  0.49$^{*}$ & 1.86 (0.25)  &  2.35  (0.24) \\
Feelings about the economy & 0.41 & 0.49 (0.25)  & 0.90 (0.23)  \\
\hline
N &  & 360 & 356 \\
\hline
\end{tabular}
}
\medskip
\vspace{2pt}
\begin{minipage}{0.99\textwidth} 
{\small     Note: Following the GoodNews and BadNews treatments, manipulation check questions were introduced to measure the effectiveness of our experimental conditions. Participants were asked to indicate their current feelings regarding their own earnings prospects, others' earnings prospects, and the overall economy using a scale ranging from $-10$ (Pessimistic) to $10$ (Optimistic), with $0$ representing Neutral feelings. Robust standard errors are presented in parentheses. One-sided t-test p-value thresholds are as follows: $^{*}$p$<$0.1; $^{**}$p$<$0.05; $^{***}$p$<$0.01.}
\end{minipage}
\end{table}
\FloatBarrier

\noindent \begin{quote}
    At the beginning of the study, we showed participants a brief ``cheap-talk'' statement to foster thoughtful and truthful responses. To gauge students' engagement levels, we included three pattern-detection questions in the demographic survey section. These questions were designed to be easily recognizable, as our aim was not to test fluid intelligence but to measure attention levels. Of the 716 subjects, 92\%, 6\%, and 2\% accurately identified three, two, and one patterns, respectively. This suggests a high level of attentiveness among the study participants.
\end{quote}



\section*{Supplementary material}

Forbes reports that, based on data from the National Center for Education Statistics, the median starting salary for college graduates is $59,600 per year.

---

What do you believe is the probability that your starting salary will exceed $59,600 after graduation?

| 0 | 10 | 20 | 30 | 40 | 50 | 60 | 70 | 80 | 90 | 100 |

Probability

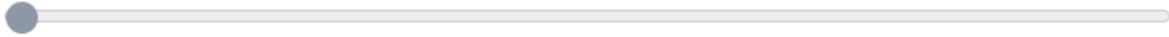

---

What do you believe is the probability that the median student with the same major as yours will have a starting salary exceeding $59,600 after graduation?

| 0 | 10 | 20 | 30 | 40 | 50 | 60 | 70 | 80 | 90 | 100 |

Probability

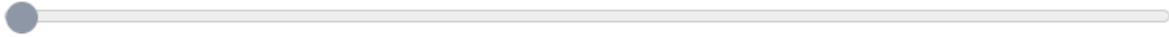

---

What is your expected starting salary after graduation in USD?

| 0 | 40000 | 80000 | 120000 | 160000 | 200000 |

Expected Salary

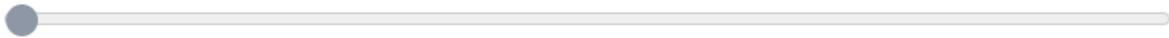

Since its launch on November 30, 2022, **ChatGPT** has generated intense discussions about how **Artificial Intelligence (AI)** may transform the labor market in the near future.

As someone majoring or planning to major in **STEM**, you might be interested in the ongoing debate about the potential impacts of ChatGPT on the labor market. On the next screen, you will find selected excerpts from a recent article published in a magazine specializing in Technology and Innovation news. This news source has a *Factual Grade* of **76%** on a 0-100% scale. Factual Grade evaluates how *well-sourced* and *informative* this source is.

Since its launch on November 30, 2022, **ChatGPT** has generated intense discussions about how **Artificial Intelligence (AI)** may transform the labor market in the near future.

As someone majoring or planning to major in **NON-STEM**, you might be interested in the ongoing debate about the potential impacts of ChatGPT on the labor market. On the next screen, you will find selected excerpts from a recent article published in a magazine specializing in Technology and Innovation news. This news source has a *Factual Grade* of **76%** on a 0-100% scale. Factual Grade evaluates how *well-sourced* and *informative* this source is.

## ChatGPT is about to revolutionize the economy. We need to decide what that looks like.

ChatGPT and other recently released generative AI technologies hold the promise of automating all sorts of tasks that were previously thought to be solely in the realm of human creativity and reasoning, from writing to creating graphics to summarizing and analyzing data. AI models are getting more powerful: they're trained on ever more data, and the number of parameters—the variables in the models that get tweaked—is rising dramatically.

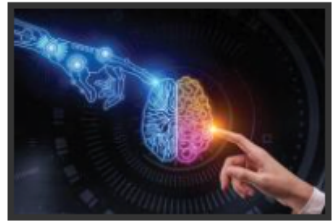

Will ChatGPT make the already troubling income and wealth inequality in the US and many other countries even worse? Could it in fact provide a much-needed boost to productivity?

> ChatGPT and similar AI models will prove to be a **powerful tool for many workers, improving their capabilities and expertise, while providing a boost to the overall economy.** Companies can quickly take up the AI tools, becoming so much more productive that they dominate their workplaces and their sectors. The AI tool may also help the least skilled and accomplished workers the most, decreasing the performance gap between employees.

**GoodNews**

# ChatGPT is about to revolutionize the economy. We need to decide what that looks like.

ChatGPT and other recently released generative AI technologies hold the promise of automating all sorts of tasks that were previously thought to be solely in the realm of human creativity and reasoning, from writing to creating graphics to summarizing and analyzing data. AI models are getting more powerful: they're trained on ever more data, and the number of parameters—the variables in the models that get tweaked—is rising dramatically.

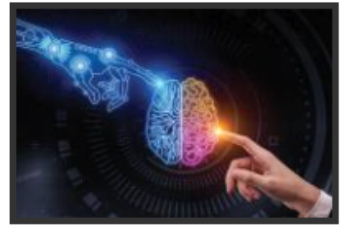

Will ChatGPT make the already troubling income and wealth inequality in the US and many other countries even worse? Could it in fact provide a much-needed boost to productivity?

> *ChatGPT and similar AI models will **destroy what once looked like automation-proof jobs, well-paying ones that require creative skills and logical reasoning; it will do little for overall economic growth.** Companies will replace relatively well-paying white-collar jobs with this new form of automation, sending those workers off to lower-paying service employment while the few who are best able to exploit the new technology reap all the benefits.*

**BadNews**

How do you feel about your future earning perspectives?

Pessimistic                                    Neutral                              Optimistic

-10        -8        -6        -4        -2        0        2        4        6        8        10

use slider to answer

---

How do you feel about the future earning perspectives of students with the same major as yours?

Pessimistic                                    Neutral                              Optimistic

-10        -8        -6        -4        -2        0        2        4        6        8        10

use slider to answer

---

How do you feel about the future growth potential of the economy?

Pessimistic                                    Neutral                              Optimistic

-10        -8        -6        -4        -2        0        2        4        6        8        10

use slider to answer

After discussing the potential impacts of ChatGPT and other AI technologies on labor market, we want to ask you these questions again.

What do you believe is the probability that your starting salary will exceed $59,600 after graduation?

| 0 | 10 | 20 | 30 | 40 | 50 | 60 | 70 | 80 | 90 | 100 |

Probability

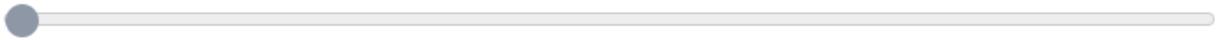

What do you believe is the probability that the median student with the same major as yours will have a starting salary exceeding $59,600 after graduation?

| 0 | 10 | 20 | 30 | 40 | 50 | 60 | 70 | 80 | 90 | 100 |

Probability

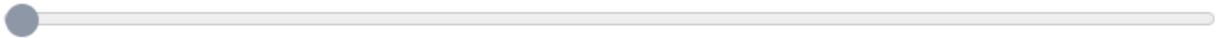

What is your expected starting salary after graduation in USD?

| 0 | 40000 | 80000 | 120000 | 160000 | 200000 |

Expected Salary

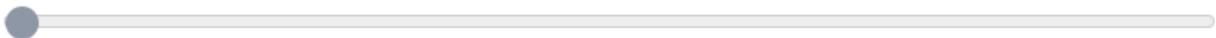



\end{document}